\documentclass[12pt, preprint]{aastex}
\usepackage{epsf}
\usepackage{psfig,epsfig}
\newcommand{\etal}{{et al}\/.}

\begin{document}
\slugcomment{Draft of \today} \shorttitle{X-ray lobes}
\shortauthors{J.H. Croston \etal} \title{An X-ray study of magnetic
field strengths and particle content in the lobes of FRII radio
sources} \author{J.H. Croston\altaffilmark{1,4}, M.J.
Hardcastle\altaffilmark{2,4}, D.E.\ Harris\altaffilmark{3}, E.
Belsole\altaffilmark{4}, M. Birkinshaw\altaffilmark{4} and D.M.
Worrall\altaffilmark{4}} \altaffiltext{1}{Service d'Astrophysique, CEA
Saclay, L'Orme des Merisiers, 91191 Gif-sur-Yvette, France.\\ Email:
jcroston@discovery.saclay.cea.fr} \altaffiltext{2}{School of Physics,
Astronomy and Mathematics, University of Hertfordshire, College Lane,
Hatfield AL10 9AB, UK} \altaffiltext{3}{Harvard-Smithsonian Center for
Astrophysics, 60 Garden Street, Cambridge, MA~02138, USA}
\altaffiltext{4}{Department of Physics, University of Bristol, Tyndall
Avenue, Bristol BS8 1TL, UK}

\begin{abstract}
We present a {\it Chandra} and {\it XMM-Newton} study of X-ray
emission from the lobes of 33 classical double radio galaxies and
quasars. We report new detections of lobe-related X-ray emission in 11
sources. Together with previous detections we find that X-ray emission
is detected from at least one radio lobe in $\sim 75$ percent of the
sample. For all of the lobe detections, we find that the measured
X-ray flux can be attributed to inverse-Compton scattering of the
cosmic microwave background radiation, with magnetic field strengths
in the lobes between (0.3 -- 1.3) $B_{eq}$, where the value $B_{eq}$
corresponds to equipartition between the electrons and magnetic field
assuming a filling factor of unity. There is a strong peak in the
magnetic field strength distribution at $B \sim 0.7 B_{eq}$. We find
that $> 70$ percent of the radio lobes are either at equipartition or
electron dominated by a small factor. The distribution of measured
magnetic field strengths differs for narrow-line and broad-line
objects, in the sense that broad-line radio galaxies and quasars
appear to be further from equipartition; however, this is likely to be
due to a combination of projection effects and worse systematic
uncertainty in the X-ray analysis for those objects. Our results
suggest that the lobes of classical double radio sources do not
contain an energetically dominant proton population, because this
would require the magnetic field energy density to be similar to the
electron energy density rather than the overall energy density in
relativistic particles.
\end{abstract}

\maketitle

\section{Introduction}
\label{intro}

Detections of X-ray inverse-Compton emission from components of radio
galaxies have the potential to resolve long-standing questions about
their particle content and magnetic field strength, because they allow
direct measurements of electron energy density, unlike observations of
radio synchrotron emission, where the electron density and magnetic
field strength cannot be decoupled. This technique has been
successfully used to measure magnetic field strengths in the hotspots
and lobes of FRII \citep{FR74} radio galaxies and quasars.
Measurements of the internal energy density in relativistic electrons
obtained from X-ray lobe detections can be used to constrain source
dynamics and particle content by allowing a comparison of the internal
pressure with the external pressure from X-ray-emitting hot gas
\citep[e.g.,][]{h02a,c04}. This is particularly important in view of
the continuing uncertainty about the dynamical status and confinement
of FRII lobes \citep{hw00a}.

There are several possible sources of photons to be inverse-Compton
scattered up to X-ray energies by the radio-synchrotron-emitting
electron population. In hotspots, where the electron density is high,
the dominant photon population comes from the radio emission itself;
this is the synchrotron-self Compton process \citep{h04b}. However, in
the lobes the density of synchrotron photons is much lower, so that
the photon energy density from the cosmic microwave background
typically dominates over that from the radio synchrotron emission. In
addition, the photon field from the nuclear source may be important in
some sources \citep{bru97}. {\it Chandra} and {\it XMM-Newton} have
allowed a number of detections of X-ray emission from lobes to be
made. In many sources the lobes have been claimed to be near to
equipartition \citep[e.g.,][]{h02a,bel04,c04,bon04,oz04} whereas in
others significant electron dominance is claimed
\citep[e.g.,][]{iso02,com03,kat03}. However, the results are dependent
on the assumed electron energy spectrum and photon population
characteristics. The unknown properties of the electron population at
energies below those observable in the radio also introduce
considerable uncertainty \citep[e.g.,][]{har04}. Differences in the
methods used to separate thermal and non-thermal X-ray emission, and
in the calculations of inverse-Compton emissivity can also be
important, so that estimates from different authors are often not
directly comparable. As yet there is no overall picture of the
magnetic field properties of the FRII population.

The particle content in radio galaxies and quasars has been the
subject of debate over several decades. There are arguments in favor
of electron-proton jets, principally based on energy transport close
to the active nucleus \citep[e.g.,][]{cf93}; however, several
independent arguments favor electron-positron jets
\citep[e.g][]{war98,hom99,kin04}. Inverse-Compton studies can provide
indirect information on the particle content on the large scale:
although relativistic protons are not directly observable by this
process, inverse-Compton observations that are consistent with
equipartition between the magnetic field and electron energy densities
make it difficult to accommodate an energetically dominant population
of protons in the lobes. \cite{h04b} have already used this argument
to suggest that hotspot energetics are not dominated by protons, and
\cite{h02a} and \cite{c04} have applied it to individual FRII lobes;
extending it to FRII lobes in general would be of considerable
interest.

In this paper we use the {\it Chandra} archives (together with some
{\it XMM-Newton} data) to compile a large sample of FRII radio
galaxies and quasars with which to investigate the inverse-Compton
properties of lobes. Our approach differs from that of \cite{kat04} in
that we do not only select sources with known hotspot, jet or lobe
emission, but include sources whose extended components have not
previously been detected in X-rays: this allows us to consider limits
on the magnetic field strength, and gives us a much larger sample (as
well as allowing us to make some new lobe detections). In
Section~\ref{sec:stat}, we compare our results with theirs. Throughout
the paper we use a cosmology in which $H_0 = 70$ km s$^{-1}$
Mpc$^{-1}$, $\Omega_{\rm m} = 0.3$ and $\Omega_\Lambda = 0.7$.
Spectral indices $\alpha$ are the energy indices and are defined in
the sense $S_{\nu} \propto \nu^{-\alpha}$.

\section{Sample and data analysis}

\subsection{Sample}

The sample was compiled from the list of 3C FRII radio sources for
which public {\it Chandra} observations existed as of early 2004,
which comprises 36 objects \citep{h04b}. We also included 4 {\it XMM-Newton}
observations of 3C radio galaxies for which detailed analysis of
lobe-related emission has previously been carried out
\citep{bel04,c04}. This gave a total sample size of 40 objects, which
span a redshift range of $\sim 0.05 - 2$.  In Table~\ref{lobe_sample}, the full
sample is listed, together with details of the observations and
references to previously published work.

\clearpage

\begin{deluxetable}{lrrrrrrrl}
\tablecaption{The sample of X-ray observed FRII radio sources.}
\tablewidth{17cm}
\tablehead{Source&$z$&Type&$\alpha_{\rm R}$&OBS&Obsid&Date observed&Duration (s)&Reference}
\startdata
3C\,6.1&0.8404&N&0.68&C&3009&2002 Oct 15&36492&2\\
3C\,9&2.012&Q&1.12&C&1595&2001 Jun 10&19883&1\\
3C\,47&0.425&Q&0.98&C&2129&2001 Jan 16&44527&2\\ 
3C\,109&0.306&B&0.85&C&4005&2003 Mar 23&45713&2\\
3C\,123&0.2177&E&0.70&C&829&2000 Mar 21&38465&14\\
3C\,173.1&0.292&E&0.88&C&3053&2002 Nov 06&23999&2\\
3C\,179&0.846&Q&0.73&C&2133&2001 Jan 15&9334&3\\ 
3C\,184&0.994&N&0.86&C&3226&2002 Sep 22&18886&4\tablenotemark{a}\\
3C\,200&0.458&N&0.84&C&838&2000 Oct 06&14660&\\
3C\,207&0.684&Q&0.90&C&2130&2000 Nov 04&37544&5\\
3C\,212&1.049&Q&0.92&C&434&2000 Oct 26&18054&15\\
3C\,215&0.411&Q&1.06&C&3054&2003 Jan 02&33803&2\\
3C\,219&0.1744&B&0.81&C&827&2000 Oct 11&17586&6\\ 
3C\,220.1&0.61&N&0.93&C&839&1999 Dec 29&18922&16\\
3C\,223&0.1368&N&0.74&X&0021740101&2003 May 30&34000&7\tablenotemark{a}\\ 
3C\,228&0.5524&N&1.0&C&2453&2001 Apr 23&13785&\\ 
3C\,254&0.734&Q&0.96&C&2209&2001 Mar 26&29668&17\\
3C\,263&0.652&Q&0.82&C&2126&2000 Oct 28&44148&8\tablenotemark{b}\\
3C\,265&0.8108&N&0.96&C&2984&2002 Apr 25&58921&9\\ 
3C\,275.1&0.557&Q&0.96&C&2096&2001 Jun 02&24757&10\\
3C\,280&0.996&N&0.81&C&2210&2001 Aug 27&63528&17\\
3C\,281&0.602&Q&0.71&C&1593&2001 May 30&15851&10\\ 
3C\,284&0.2394&N&0.95&X&0021740201&2002 Dec 12&43000&7\tablenotemark{a}\\
3C\,292&0.710&N&0.80&X&0147540101&2002 Oct 29&34000&4\tablenotemark{a}\\ 
3C\,294&1.78&N&1.07&C&3207&2002 Feb 27&122020&11\tablenotemark{c}\\
3C\,295&0.4614&N&0.63&C&2254&2001 May 18&90936&18\\
3C\,303&0.141&B&0.76&C&1623&2001 Mar 23&14951&19\\
3C\,321&0.096&N&0.60&C&3138&2002 Apr 30&47130&2\\
3C\,322&1.681&N&0.81&X&0028540301&2002 May 17&43000&4\tablenotemark{a}\\
3C\,324&1.2063&N&0.90&C&326&2000 Jun 25&42147&2\\
3C\,330&0.5490&N&0.71&C&2127&2001 Oct 16&44083&8\tablenotemark{b}\\ 
3C\,334&0.555&Q&0.86&C&2097&2001 Aug 22&32468&\\ 
3C\,351&0.371&Q&0.73&C&2128&2001 Aug 24&45701&8\tablenotemark{b}\\
3C\,390.3&0.0569&B&0.75&C&830&2000 Apr 17&33974&\\
3C\,401&0.201&E&0.71&C&3083&2002 Jul 20&22666&20\\
3C\,403&0.0590&N&0.74&C&2968&2002 Dec 07&49472&12\tablenotemark{a}\\
3C\,405&0.0590&N&0.74&C&360&2000 May 21&34720&21\\
3C\,427.1&0.572&E&0.97&C&2194&2002 Jan 27&39456&\\ 
3C\,438&0.290&E&0.88&C&3967&2002 Dec 27&47272&\\
3C\,452&0.0811&N&0.78&C&2195&2001 Aug 21&79922&13\\ 
\enddata
\label{lobe_sample}
\tablecomments{Column 3 gives the radio-source type: N is a
  narrow-line radio galaxy, B is a broad-line radio galaxy, E is a
  low-excitation radio galaxy, and Q is a radio-loud quasar. Column 4
  gives the low-frequency radio spectral index (normally between 178
  and 750 MHz). Column 5 indicates whether the observation was with
  {\it Chandra} (C) or {\it XMM-Newton} (X). Column 6 gives the {\it
  Chandra} or {\it XMM-Newton} observing ID. Columns 7 and 8 give the
  observing date and original livetime. References for observations already
  discussed in the literature are in column 9.} \tablerefs{(1)
  \citealt{fcj03}; (2) \citealt{h04b}; (3) \citealt{sam02}; (4)
  \citealt{bel04}; (5) \citealt{brun02}; (6) \citealt{com03}; (7)
  \citealt{c04}; (8) \citealt{h02a}; (9) \citet{bon03}; (10)
  \citealt{cf03}; (11) \citealt{cf03b}; (12) \citet{kra05}; (13)
  \citealt{iso02}; (14) \citealt{hbw01}; (15) \citealt{ald03};
  (16) \citealt{wor01}; (17) \citealt{don03}; (18) \citealt{bcs01};
  (19) \citealt{kat03b}; (20) \citealt{rey04}; (21) \citealt{wil00}.} \tablenotetext{a}{The authors' analysis methods are
  similar to those used in the current paper, so we do not reanalyse
  this observation.} \tablenotetext{b}{The authors' analysis methods
  are similar to those used in the current paper, but we repeat their
  inverse-Compton calculations to take into account a different
  cosmology and low-energy electron cutoff.} \tablenotetext{c}{The
  authors argue that there is IC emission produced by scattering of
  CMB photons around this high redshift source; however, the emission
  is not coincident with the lobes or any detected radio emission, so
  this is not an X-ray lobe detection by our definition.}
\end{deluxetable}

\clearpage

\subsection{X-ray analysis}

Nine of the sources in the sample have previous detections of
lobe-related emission analysed by workers including a subset of the
present authors, with analysis based around the {\sc synch} code of
\cite{hbw98}. Of the remaining objects, six have previously published
detections of lobe-related X-ray emission. For consistency with the
rest of our sample, we reanalysed the data for those lobe detections
that had not previously been analysed using our code.

We extracted the {\it Chandra} archive data for the 31 objects not
previously studied by our group, and prepared and analysed the data
using standard methods using {\sc ciao} 3.1 and {\sc caldb} 2.28. The
data were filtered for good time intervals, and an image in the energy
range 0.5 -- 5.0 keV was made for each source. We examined the X-ray
image to determine whether it would be possible to make a measurement
of the lobe-related X-ray emission. In 7 cases, we decided that it
would be impossible to determine accurately the level of X-ray
emission associated with the radio lobes, either because of
background-subtraction difficulties or other confusing components of
X-ray emission. These cases were 3C\,123, 3C\,295, 3C\,401, 3C\,405
and 3C\,438, which are all in rich clusters that in several cases show
complex structure on the scales of the radio lobes; 3C\,294, where
extended non-thermal X-ray emission, not associated with the lobes, is
present \citep{cf03b}; and 3C\,324, where hotspot-lobe separation
would be difficult \citep{h04b}. In addition, there were several
sources where it was only possible to determine the level of
lobe-related X-ray emission accurately for one of the two lobes,
because of confusion with bright AGN, jet or hotspot emission. The
sources for which only one lobe could be studied are 3C\,9, 3C\,207,
3C\,212, 3C\,254, 3C\,280, 3C\,303, and 3C\,321. In total, the
analysis was carried out for 24 radio galaxies and quasars, which,
together with the 9 previously studied objects, gives a total sample
of 54 lobes.

Spectral extraction regions were chosen based on the extent of the
radio structure, as determined using the 1.4-GHz radio maps (see the
following Section). The regions are circular, rectangular or
elliptical approximations to the extent of low-frequency radio
emission. In all but two cases spectra were obtained separately for
each radio lobe. Low-frequency radio maps were used as they best
represent the distribution of the electrons responsible for scattering
CMB photons to X-ray energies. In most cases the lobe-related emission
is of extremely low surface brightness, so that accurate background
subtraction is crucial. In many cases there is also X-ray emission
from a bright AGN and from the hot-gas environment. We therefore used
local background regions at the same distance from the core as the
source regions, so as to minimize the contamination, which is expected
to be symmetrically distributed, and in particularly difficult cases
(strong nuclear X-ray emission and small lobes) used PSF modelling
based on {\sc Chart}\footnote{See http://cxc.harvard.edu/chart/ .} and
{\sc marx} to verify the correctness of our extraction regions.
Spectral extraction was carried out using the {\sc ciao} task {\sc
psextract}, which produces source and background spectra and response
files (we compared our results with those obtained using weighted
response files generated using the {\sc acisspec} script for several
representative sources and found no significant differences; we
therefore chose to use the considerably faster {\sc psextract}
script). We used the energy range 0.5 -- 5.0 keV for spectral
analysis.

Where there were sufficient counts, the spectrum was grouped to a
minimum of 20 counts per background-subtracted bin, and a power-law
model was fitted to the extracted spectrum using {\sc xspec}. In all
cases Galactic $N_{\rm H}$ was assumed, using values from pointed
observations or obtained with the $N_{\rm H}$ tool provided by NASA's
High Energy Astrophysics Science Archive Research Center
(HEASARC)\footnote{http://heasarc.gsfc.nasa.gov/}, based on the
measurements of \citet{dl90}. In the case of 3C\,452, where
subtraction of any thermal emission was difficult due to the
radio-source morphology, we also fitted a two-component {\it mekal}
plus power-law model. In the majority of cases, however, the total
number of background-subtracted counts was too low to fit a spectrum
or there was no detected X-ray emission above the 3$\sigma$ background
level. In those cases, we assumed a power-law photon spectrum with
$\Gamma = 1.5$, and used the measured count rate, or upper limit from
the 3$\sigma$ background level, to determine the 1-keV flux density.
Table~\ref{lobespec} gives details of the spectral fits for those
sources where it was possible to fit the spectrum. In most cases an
acceptable fit is obtained for $\Gamma = 1.5$, as expected for
inverse-Compton emission by radio-synchrotron-emitting electrons with
the typical injection energy spectrum predicted from shock
acceleration \citep[e.g.][]{bell78}. Table~\ref{lobeflux} gives the
absorbing column density, number of counts and flux density
measurements for the other detected sources, and Table~\ref{nondets}
gives upper limits for the remaining sources.

\clearpage

\begin{deluxetable}{lrrrrrr}
\tablecaption{Spectral fits for X-ray
  lobe detections with sufficient counts.}
\tablewidth{15cm}
\tablehead{Source&Net counts&$N_{\rm H}$ (cm$^{-2})$&$\Gamma$&$S_{\rm 1\ keV}$ (nJy)&$\chi^{2}$/d.o.f.}
\startdata
3C\,47N&197&$5.87 \times 10^{20}$&$1.4\pm0.4$&$3.6\pm0.7$&4.9/6\\
3C\,47S&434&$5.87 \times 10^{20}$&$1.9\pm0.2$&$10\pm1$&21/15\\
3C\,215N&109&$3.75 \times 10^{20}$&$1.4\pm0.3$&$2.9\pm0.4$&1/3\\
3C\,215S&119&$3.75 \times 10^{20}$&$1.5\pm0.5$&$2.9\pm0.5$&2.9/3\\
3C\,219N&188&$1.51 \times 10^{20}$&$2.0\pm0.3$&$9\pm1$&3.6/6\\
3C\,219S&147&$1.51 \times 10^{20}$&$1.7\pm0.5$&$7\pm1$&7/4\\
3C\,265E&142&$1.90 \times 10^{20}$&$1.9\pm0.2$&$3.1\pm0.3$&1/5\\
3C\,452 (Model I)&2746&$1.19 \times 10^{21}$&$1.75\pm0.09$&$37\pm2$&96/89\\
3C\,452 (Model II)&2746&$1.19 \times 10^{21}$&$1.5$ (frozen)&$23\pm4$&87/88\\
\enddata
\label{lobespec}
\tablecomments{Spectra were fitted in the energy range 0.5 -- 5.0 keV.
  Column 2 gives the {\it Chandra} background-subtracted 0.5 -- 5.0
  keV counts in the lobe. Column 3 gives the assumed Galactic hydrogen
  column density, frozen for the purposes of the fit. Errors in
  columns 4 and 5 are the statistical errors, $1\sigma$ for one
  interesting parameter. Two models were fitted to the 3C\,452 data,
  as described in the text. Model II includes a thermal component wiht
  $kT = 0.6\pm0.3$ keV, consistent with the results of
  \citealt{iso02}.}
\end{deluxetable}

\clearpage

\begin{deluxetable}{lrrr}
\tablecaption{X-ray flux measurements for detected
  lobes with insufficient counts for spectral fitting}
\tablewidth{10cm}
\tablehead{Source&Net counts&$N_{\rm H}$ (cm$^{-2}$)&$S_{\rm 1\ keV}$ (nJy)}
\startdata
3C\,9W&13&$4.11 \times 10^{20}$&$0.6\pm0.3$\\
3C\,109N&70&$1.57 \times 10^{21}$&$1.5\pm0.3$\\
3C\,109S&69&$1.57 \times 10^{21}$&$1.5\pm0.4$\\
3C\,173.1N&17&$5.25 \times 10^{20}$&$0.6\pm0.2$\\
3C\,179E&17&$4.31 \times 10^{20}$&$1.3\pm0.4$\\
3C\,179W&9&$4.31 \times 10^{20}$&$0.7\pm0.3$\\
3C\,200&35&$3.69 \times 10^{20}$&$1.6\pm0.4$\\
3C\,207W&23&$5.40 \times 10^{20}$&$0.6\pm0.2$\\
3C\,265W&46&$1.90 \times 10^{20}$&$0.7\pm0.2$\\
3C\,275.1S&20&$1.89 \times 10^{20}$&$0.5\pm0.1$\\
3C\,280W&18&$1.25 \times 10^{20}$&$0.2\pm0.1$\\
3C\,281N&25&$2.21 \times 10^{20}$&$1.0\pm0.3$\\
3C\,334N&36&$4.24 \times 10^{20}$&$0.9\pm0.3$\\
3C\,334S&36&$4.24 \times 10^{20}$&$0.9\pm0.2$\\
3C\,427.1S&14&$1.09 \times 10^{21}$&$0.3\pm0.1$\\
\enddata
\label{lobeflux}
\tablecomments{Column 2 gives the {\it Chandra} background-subtracted
  0.5 -- 5.0 keV counts in the lobe. The 1-keV flux densities were
  determined by assuming a power-law with $\Gamma = 1.5$, as described
  in the text.}
\end{deluxetable}

\clearpage

\begin{deluxetable}{lrrr}
\tablecaption{Upper limits
  on the unabsorbed 1-keV flux density for the non-detected lobes.}
\tablewidth{10cm}
\tablehead{Source&Net counts&$N_{\rm H}$ (cm$^{-2}$)&$S_{\rm 1\ keV}$ (nJy)}
\startdata
3C\,6.1N&$<14$&$1.75 \times 10^{21}$&$<0.4$\\
3C\,6.1S&$<15$&$1.75 \times 10^{21}$&$<0.5$\\
3C\,173.1S&$<17$&$5.25 \times 10^{20}$&$<0.6$\\
3C\,212S&$<42$&$4.09 \times 10^{20}$&$<1.7$\\
3C\,220.1N&$<40$&$1.93 \times 10^{20}$&$<1.2$\\
3C\,220.1S&$<35$&$1.93 \times 10^{20}$&$<1.1$\\
3C\,228N&$<12$&$3.18 \times 10^{20}$&$<0.8$\\
3C\,228S&$<11$&$3.18 \times 10^{20}$&$<0.7$\\
3C\,254W&$<16$&$1.75 \times 10^{20}$&$<0.4$\\
3C\,275.1N&$<11$&$1.89 \times 10^{20}$&$<0.3$\\
3C\,281S&$<20$&$2.21 \times 10^{20}$&$<0.8$\\
3C\,303E&$<23$&$1.60 \times 10^{20}$&$<1.0$\\
3C\,321W&$<43$&$4.10 \times 10^{20}$&$<0.7$\\
3C\,390.3N&$<86$&$3.74 \times 10^{20}$&$<1.8$\\
3C\,390.3S&$<124$&$3.74 \times 10^{20}$&$<2.7$\\
3C\,427.1N&$<16$&$1.09 \times 10^{21}$&$<0.4$\\
\enddata
\label{nondets}
\tablecomments{Column 2 gives the $3\sigma$ upper limit of {\it
    Chandra} background-subtracted 0.5 -- 5.0 keV counts. The upper
  limit 1-keV flux densities were determined by assuming a power-law
  model with $\Gamma = 1.5$, as described in the text.}
\end{deluxetable}

\clearpage

\begin{figure}
\begin{center}
\plotone{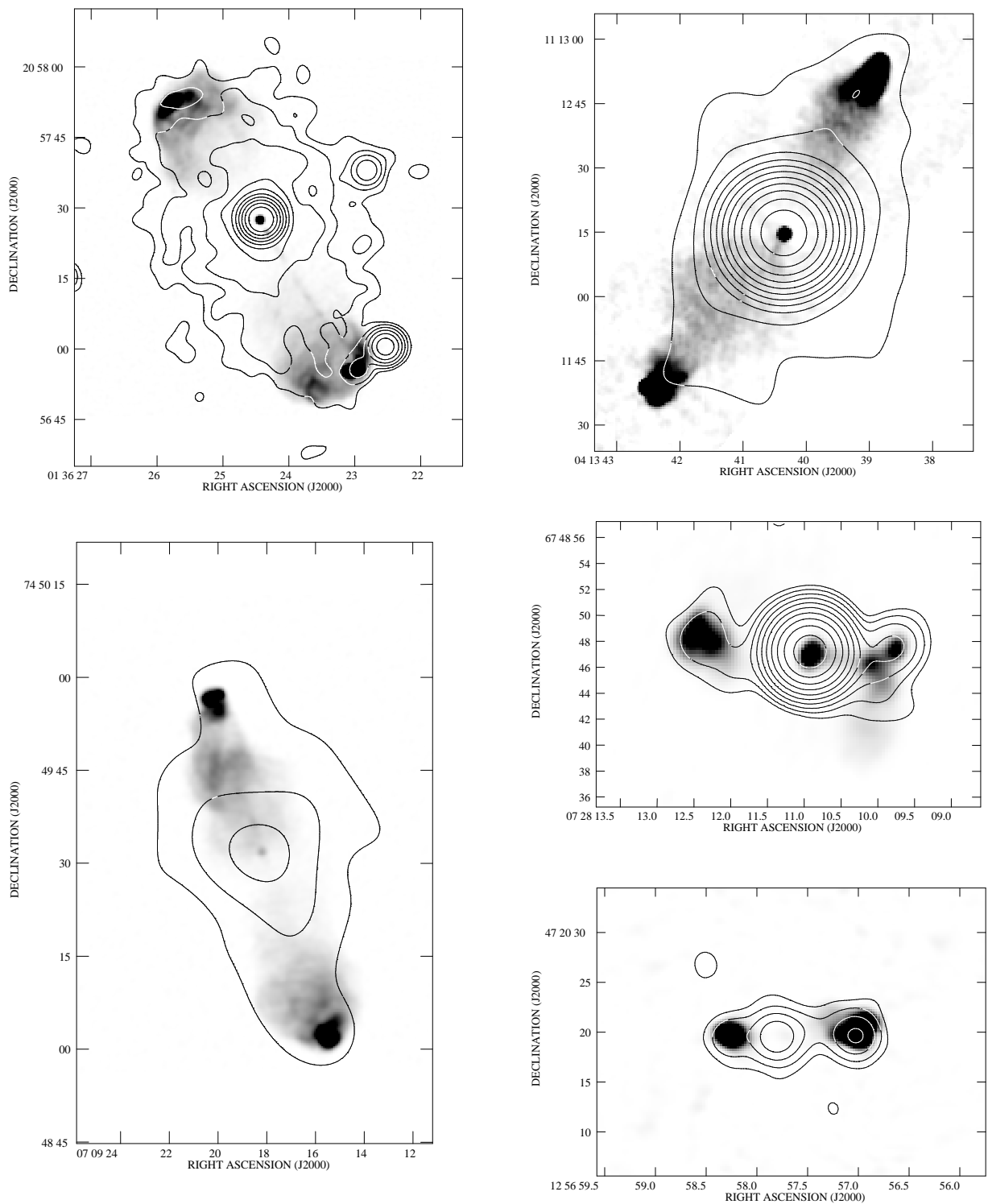}
\caption[{\it Chandra} images of the X-ray emission from 3C\,47,
  3C\,109, 3C\,179, 3C\,280, and 3C\,173.1]{Contour maps from Gaussian
  smoothed 0.5 -- 5.0 keV {\it Chandra} images of the X-ray emission
  from (clockwise from top left): 3C\,47 ($\sigma = 1.7$ arcsec),
  3C\,109 ($\sigma = 4.9$ arcsec), 3C\,179 ($\sigma = 1.2$ arcsec),
  3C\,280 ($\sigma = 1.2$ arcsec), and 3C\,173.1 ($\sigma = 4.4$
  arcsec). The X-ray contour levels are at 1,2,4 ... $\times 3\sigma$
  level, calculated using the method of \cite{hwb98}. Radio maps shown in
  grayscale are from the 1.4-GHz radio maps listed in
  Table~\ref{loberadio}.}
\label{lobes1}
\end{center}
\end{figure}

\begin{figure}
\begin{center}
\epsscale{.80}
\plotone{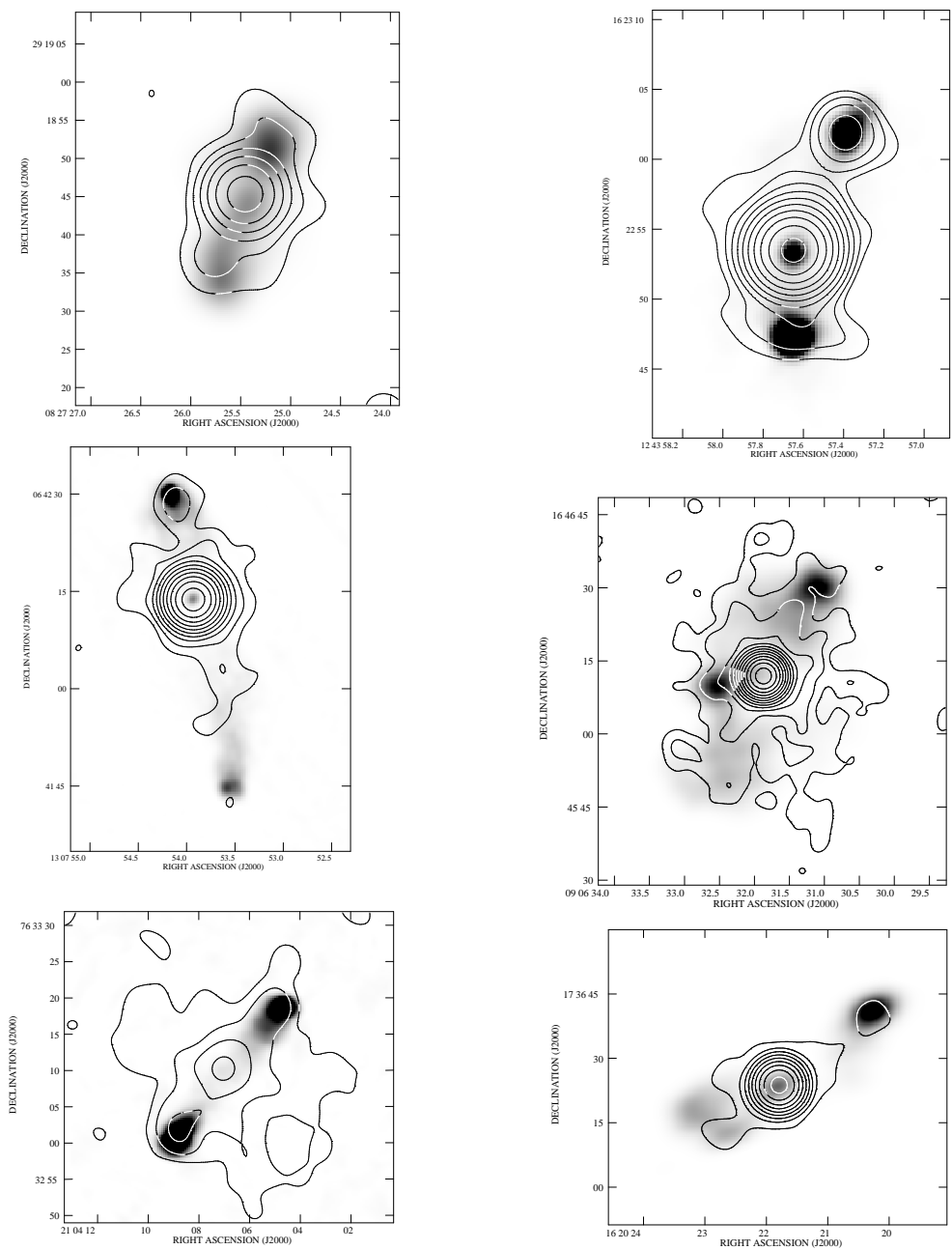}
\caption[{\it Chandra} images of the X-ray emission from 3C\,200,
  3C\,215, 3C\,275.1, and 3C\,228]{Contour maps from Gaussian smoothed
  0.5 -- 5.0 keV {\it Chandra} images of the X-ray emission from
  (clockwise from top left): 3C\,200 ($\sigma = 2.4$ arcsec),
  3C\,275.1 ($\sigma = 1.2$ arcsec), 3C\,215 ($\sigma = 1.7$ arcsec),
  3C\,334 ($\sigma = 2.4$ arcsec), 3C\,427.1 ($\sigma = 1.7$ arcsec)
  and 3C\,281 ($\sigma = 1.7$ arcsec). The X-ray contour levels are at
  1,2,4 ... $\times 3\sigma$ level, calculated using the method of
  \cite{hwb98}. Radio maps shown in grayscale are from the 1.4-GHz
  radio maps listed in Table~\ref{loberadio}.}
\label{lobes2}
\end{center}
\end{figure}

In total, of the 39 lobes analysed, 23 were detected at the $3\sigma$
level or above. In Figs.~\ref{lobes1} and \ref{lobes2} we show contour
maps made from smoothed images for each of the sources with at least
one lobe detection where the data are unpublished, or where the lobe
detection has not previously been presented (3C\,47, 3C\,109,
3C\,173.1, 3C\,179, 3C\,200, 3C\,215, 3C\,275.1, 3C\,280, 3C\,281,
3C\,334, and 3C\,427.1). Radio maps are shown in grayscale to
illustrate the relation between radio and X-ray emission.

\clearpage

\clearpage

\clearpage

\subsection{Radio data}
\label{sec:rdata}

The electrons responsible for scattering CMB photons to X-ray energies
have $\Gamma \sim 1000$, and so their radio emission is emitted at
$\nu \sim 100$ MHz, assuming a typical magnetic field strength of 1.5
nT (1nT = 10$\mu$G). This is at the lower end of the observable
radio region. It is therefore essential to use the lowest-frequency
radio maps available that have sufficient resolution to determine the
extent of the radio emission and the radio spectrum of the lobes. We
used the 178-MHz flux densities from the 3C and 3CRR catalogues, and
obtained 1.4-GHz flux densities for each lobe using the best available
maps.

For those sources where we did not have access to a 1.4-GHz radio map,
we extracted archive VLA data (choosing a VLA configuration that
samples the largest angular structure of each source, so as to include
all of the source flux) and performed calibration and mapping using
standard techniques in {\sc aips}. Table~\ref{loberadio} lists the
radio maps used to determine the ratio of lobe flux densities and to
define the X-ray spectral extraction regions.

1.4-GHz flux densities were measured using {\sc tvstat} in AIPS. The
entire extent of low-frequency radio emission was measured for each
lobe, as the X-ray extraction regions were chosen using the same
maps. The flux from any hotspots or jets was excluded. 178-MHz flux
densities for each lobe were then estimated by scaling the 3C or 3CRR
flux densities based on the ratio between the 1.4-GHz flux densities
for that lobe and the total 1.4-GHz flux densities from the
lobes. Here we assume that the 178-MHz flux density is dominated by
emission from the radio lobes, so that jet and hotspot emission is not
important at that frequency. This procedure also implicitly assumes
that the low-frequency spectral indices are the same for both lobes of
a given source. In general this assumption has not been tested, but
since we know the high-frequency spectral indices of the lobes in a
given source are rarely very different \citep{liup91}, it seems
unlikely that it is seriously wrong. We have verified that
low-frequency lobe spectral indices are similar where suitable data
(e.g., 330-MHz radio maps) are available to us: the results of this
investigation suggest that the inferred 178-MHz lobe flux densities
are likely to be wrong by at most 20\%, which would correspond to a
systematic error in the predicted inverse-Compton emission of around
10\%.

\clearpage

\begin{deluxetable}{lrrr}
\tablecaption{Radio maps used in the analysis}
\tablewidth{13cm}
\tablehead{Source&Frequency (GHz)&Date observed&Reference/Proposal ID}
\startdata
3C\,6.1&1.48&1987 Oct 03&AH291\\
3C\,9&1.54&1992 Dec 13&AL280\\ 
3C\,47&1.65&&1\\
3C\,109&1.45&&2\\
3C\,173.1&1.48&&3\\
3C\,179&1.65&1986 Mar 21&AC150\\
3C\,200&1.49&1987 Nov 15&AH271\\
3C\,207&1.54&1992 Dec 13&AL280\\
3C\,212&1.66&1982 Mar 01&LAIN\\
3C\,215&1.49&1987 Nov 15&AH271\\
3C\,219&1.52&&4\\
3C\,220.1&1.40&1995 Oct 00&AH568\\
3C\,228&1.42&1986 Jul 12&AL113\\
3C\,254&1.56&1989 Feb 01&AB522\\
3C\,265&1.42&1986 Jul 12&AL113\\
3C\,275.1&1.49&1989 Jan 21&AP158\\
3C\,280&1.56&1989 Feb 01&AB522\\
3C\,281&1.43&1992 Nov 18&AB631\\
3C\,303&1.45&&3\\
3C\,321&1.51&1986 Dec 02&AV127\\
3C\,334&1.49&1986 Jul 12&AL113\\
3C\,390.3&1.45&&5\\
3C\,427.1&1.54&1986 Jun 04&AL113\\
3C\,452&1.41&&1\\
\enddata
\label{loberadio}
\tablecomments{Observation dates are given for the archive data.
  References are given for published maps (where the electronic image
  was obtained from the 3CRR database, \citealt{lea98}), and VLA
  proposal IDs for archive data.}
\tablerefs{(1) \citealt{lea98}; (2) \citealt{gio94}; (3)
  \citealt{lp91}; (4) \citealt{cla92}; (5) \citealt{lp95}.}
\end{deluxetable}

\clearpage

\section{Synchrotron and inverse-Compton modelling}
\label{sec:synchmod}

We used the X-ray flux densities or upper limits given in
Tables~\ref{lobespec}, \ref{lobeflux} and \ref{nondets}, and radio
flux densities at 178 MHz and 1.4 GHz obtained as described in
Section~\ref{sec:rdata}, to carry out synchrotron and inverse-Compton
modelling using {\sc synch} \citep{hbw98} for the sources not
previously analysed using this method. The radio lobes were modelled
either as spheres, cylinders or prolate ellipsoids, depending on the
morphology of the low-frequency radio emission. As the angle to the
line of sight is not well constrained for most of the sources, the
source dimensions are the {\it projected} dimensions. This is not a
good approximation, as sources in the sample will lie at all angles to
the line of sight. We discuss the likely effects of this approximation
later.

In each case we used the radio flux densities to normalize the
synchrotron spectrum. We initially assumed a broken power-law electron
distribution with initial electron energy index, $\delta$, of 2,
$\gamma_{{\rm min}} = 10$ and $\gamma_{{\rm max}} = 10^{5}$, and a
break energy in the range $\gamma_{{\rm break}} = 1200$ -- $10000$,
chosen so as to fit the two radio data points. In many cases the
assumed spectral break (of 1 in electron energy index) was not
sufficiently large to fit the radio data. In these cases we instead
lowered $\gamma_{{\rm max}}$ to fit the high-frequency slope of the
radio spectrum. The effective $\gamma_{{\rm max}}$ is expected to
decrease as the synchrotron plasma ages and/or expands, so this is a
physically plausible change to make. The choice of $\gamma_{{\rm
max}}$ does not significantly affect the prediction for CMB IC, as
electrons with $\gamma \ll \gamma_{{\rm max}}$ are responsible for the
scattering to X-ray energies. (The prediction for SSC emission is
significantly reduced if $\gamma_{{\rm max}}$ is reduced; however, SSC
is not the dominant emission process in any of the sources.) The
choice of parameters that affect the low-energy electron population
($\delta$ and $\gamma_{{\rm min}}$) has a more important effect on the
predicted inverse-Compton flux. We therefore discuss the effect of
modifying these parameters on our results and justify our adopted
values in more detail in Section~\ref{sec:lowfreq}. In
Table~\ref{synchmod}, we give the parameters of the synchrotron model
for each radio lobe.

We then determined the predictions for CMB IC and SSC at 1 keV based
on the modelled synchrotron spectrum for each source, assuming
equipartition between radiating particles and magnetic field and a
filling factor of unity. For 3C\,452, we adopted the 1-keV flux
density from the two-component fit, which gave a better fit statistic
than the single power-law model. Table~\ref{icresults} gives the
observed and predicted fluxes for each source in the sample, including
the previously published sources.

\clearpage

\begin{deluxetable}{lrrrrrrr}
\tablecaption{Synchrotron model parameters and radio spectral
  information for each source.}
\tablehead{
Source&$\gamma_{\rm max}$&$\gamma_{\rm break}$&Shape&$r$&$S_{178}$&$S_{1.4}$\\
&&&&(arcsec)&(Jy)&(Jy)}
\tablewidth{11cm}
\startdata
3C\,6.1N&3000&&C&4.96&9.7&0.37\\
3C\,6.1S&3000&&C&4.22&5.2&0.2\\
3C\,9W&3000&&S&2.7&14.6&0.67\\
3C\,47N&6000&&S&15&13.1&0.69\\
3C\,47S&6000&&S&17.96&15.7&0.83\\
3C\,109N&4000&&E&13&11.0&0.86\\
3C\,109S&5000&&E&13.6&12.5&0.98\\
3C\,173.1N&5000&&E&8.34&9.4&0.83\\
3C\,173.1S&6000&&S&10.5&7.4&0.66\\
3C\,179E&100000&3000&S&5.29&6.7&0.94\\
3C\,179W&100000&3000&S&3.81&2.6&0.37\\
3C\,200&100000&2000&S&13.3&12.3&1.52\\
3C\,207W&3000&&S&4.32&5.9&0.24\\
3C\,212S&4000&&S&3&0.07\tablenotemark{a}&0.65\\
3C\,215N&5000&&S&13.5&7.3&0.41\\
3C\,215S&6000&&E&13.4&5.1&0.29\\
3C\,219N&100000&4000&C&45.3&23.3&2.9\\
3C\,219S&100000&3400&C&34.5&21.6&2.7\\
3C\,220.1E&100000&1200&S&11.1&8.0&0.86\\
3C\,220.1W&100000&1200&S&9.7&9.2&1.0\\
3C\,228N&100000&2000&C&7.02&9.6&1.3\\
3C\,228S&100000&2000&C&7.36&14.2&1.9\\
3C\,254W&2600&&S&4.7&13.9&0.16\\
3C\,265E&4000&&C&9.3&11.9&0.33\\
3C\,265W&5000&&C&7.9&9.4&0.26\\
3C\,275.1N&2800&&S&6.1&8.2&0.16\\
3C\,275.1S&2800&&S&5.35&11.7&0.23\\
3C\,280W&2000&&S&3.8&14.9&0.11\\
3C\,281N&100000&1800&S&8.86&2.9&0.31\\
3C\,281S&100000&1800&C&9.68&3.11&0.34\\
3C\,303E&3000&&S&9.3&12.2&0.39\\
3C\,321W&5000&&S&33.1&7.6&0.09\\
3C\,334N&100000&2000&S&10.6&6.9&0.80\\
3C\,334S&1000&1800&S&7.88&5.0&0.50\\
3C\,390.3N&6000&&S&39&21.6&1.65\\
3C\,390.3S&6000&&S&48.3&30.2&2.3\\
3C\,427.1N&100000&2000&C&4.85&12.8&1.7\\
3C\,427.1S&100000&2000&C&4.59&16.2&2.1\\
3C\,452&100000&6000&C&88.96&59.3&10.5\\
\enddata
\label{synchmod}
\tablecomments{Shapes are S: sphere, C: cylinder and E: ellipsoid. $r$
  is the equivalent spherical radius of the modelled volume. $S_{\rm
  178}$ and $S_{1.4}$ are the assumed 178-MHz radio flux density and
  the measured 1.4-GHz flux density, respectively. $\gamma_{{\rm
  min}}$ is 10 in all cases.} \tablenotetext{a}{For this source it was
  not possible to use a 178-MHz flux density, because it was
  impossible to determine the flux ratio of the two lobes from the
  1.4-GHz map (of very low resolution). We therefore used the 8-GHz
  flux density (given here) to constrain the spectrum instead.}
\end{deluxetable}

\clearpage

\begin{deluxetable}{lrrrrr}
\tablecaption{Observed and predicted X-ray flux densities at 1 keV for the
IC models.}
\tablehead{Source&\multicolumn{4}{c}{1-keV flux densities (nJy)}&$R$\\
&Observed&Predicted SSC&Predicted
  CMB-IC&Total predicted}
\startdata
3C\,6.1N&$<0.4$&--&0.5&0.5&$<0.8$\\
3C\,6.1S&$<0.5$&--&0.3&0.3&$<1.7$\\
3C\,9W&0.6&--&0.7&0.7&$0.9 \pm 0.4$\\
3C\,47N&3.6&--&1.1&1.1&$3.3 \pm 0.6$\\
3C\,47S&10&--&1.6&1.6&$6.3 \pm 0.6$\\
3C\,109N&1.5&--&0.7&0.7&$2.1 \pm 0.4$\\
3C\,109S&1.5&--&0.8&0.8&$1.9 \pm 0.5$\\
3C\,173.1N&0.6&--&0.3&0.3&$2.0 \pm 0.7$\\
3C\,173.1S&$<0.6$&--&0.4&0.4&$<1.5$\\
3C\,179E&1.3&0.04&0.4&0.4&$3.3 \pm 1$\\
3C\,179W&0.7&0.01&0.14&0.15&$4.7 \pm 2$\\
3C\,184N&0.2&0.103&0.051&0.154&$1.3 \pm 0.7$\\
3C\,200&1.6&0.02&1.2&1.2&$1.3 \pm 0.3$\\
3C\,207W&0.6&--&0.2&0.2&$3 \pm 1$\\
3C\,212S&$<1.7$&0.006&0.05&0.06&$<28$\\
3C\,215N&2.9&--&0.7&0.7&$4.1 \pm 0.6$\\
3C\,215S&2.9&--&0.6&0.6&$4.8 \pm 0.8$\\
3C\,219N&9&0.02&3.3&3.3&$2.7 \pm 0.3$\\
3C\,219S&7&0.02&2.2&2.2&$3.2 \pm 0.2$\\
3C\,220.1E&$<1.2$&0.01&1.0&1.0&$<1.2$\\
3C\,220.1W&$<1.1$&0.02&0.9&0.9&$<1.2$\\
3C\,223N&3.1&0.004&1.3&1.4&$2.2 \pm 0.4$\\
3C\,223S&3.0&0.004&1.2&1.2&$2.5 \pm 0.4$\\
3C\,228N&$<0.8$&0.02&0.5&0.5&$<1.6$\\
3C\,228S&$<0.7$&0.04&0.6&0.6&$<1.2$\\
3C\,254W&$<0.4$&--&0.5&0.5&$<0.8$\\
3C\,263NW&0.8&0.004&0.2&0.2&$4.0 \pm 1.0$\\
3C\,263SE&0.5&0.001&0.1&0.1&$5.0 \pm 2.0$\\
3C\,265E&3.1&--&1.2&1.2&$2.6 \pm 0.3$\\
3C\,265W&0.7&--&0.8&0.8&$0.9 \pm 0.2$\\
3C\,275.1N&$<0.3$&--&0.4&0.4&$<0.8$\\
3C\,275.1S&0.5&--&0.4&0.4&$1.3 \pm 0.3$\\
3C\,280W&0.2&--&0.6&0.6&$0.3 \pm 0.2$\\
3C\,281N&1.0&0.003&0.4&0.4&$2.5 \pm 0.8$\\
3C\,281S&$<0.8$&0.003&0.5&0.5&$<1.6$\\
3C\,284E&1.9&0.003&0.93&0.94&$2.0 \pm 0.2$\\
3C\,284W&0.90&0.002&0.82&0.82&$1.1 \pm 0.2$\\
3C\,292&4.1&0.01&2.42&2.43&1.7\\
3C\,303E&$<1.0$&--&0.3&0.3&$<3.3$\\
3C\,321W&$<0.7$&--&1.0&1.0&$<0.7$\\
3C\,322&1.4&0.05&1.3&1.4&1.0\\
3C\,330NE&0.28&0.01&0.12&0.13&$2.2 \pm 0.8$\\
3C\,330SW&0.32&0.01&0.13&0.14&$2.3 \pm 0.6$\\
3C\,334N&0.9&0.01&0.7&0.7&$1.3 \pm 0.3$\\
3C\,334S&0.9&0.007&0.40&0.40&$2.3 \pm 0.5$\\
3C\,351N&1.1&0.001&0.15&0.15&$7.3 \pm 2.0$\\
3C\,351S&0.7&0.001&0.12&0.12&$5.8 \pm 2.5$\\
3C\,390.3N&$<1.8$&--&1.4&1.4&$<1.3$\\
3C\,390.3S&$<2.7$&--&2.3&2.3&$<1.2$\\
3C\,403E&1.63&0.002&0.35&0.35&$4.6 \pm 2.1$\\
3C\,403W&1.38&0.002&0.34&0.34&$4.0 \pm 2.1$\\
3C\,427.1N&$<0.4$&0.05&0.3&0.4&$<1.0$\\
3C\,427.1S&0.3&0.08&0.3&0.4&$0.8 \pm 0.3$\\
3C\,452&23&0.4&7.8&7.8&$2.9 \pm 0.5$\\
\enddata
\label{icresults}
\tablecomments{Flux densities are predicted from the radio data on the
  assumption of equipartition using the {\sc synch} code as described
  in the text. Where no SSC flux density is quoted, the predicted
  value was less than 1 pJy, and so is irrelevant to the total
  inverse-Compton flux density. $R$ is the ratio of observed to
  predicted total 1-keV flux density. Errors on the $R$ value are
  entirely due to the uncertainties on the 1-keV flux densities, as
  quoted in this paper or the papers in which they were originally
  measured, and do not take into account any systematic uncertainties.
  For the sources where the spectral modelling details are not given
  in Table~\ref{synchmod}, the results are taken from the paper
  referred to in Table~\ref{lobe_sample}.}
\end{deluxetable}

\clearpage

\section{Results}

\label{sec:stat}

\begin{figure}
\begin{center}
\plottwo{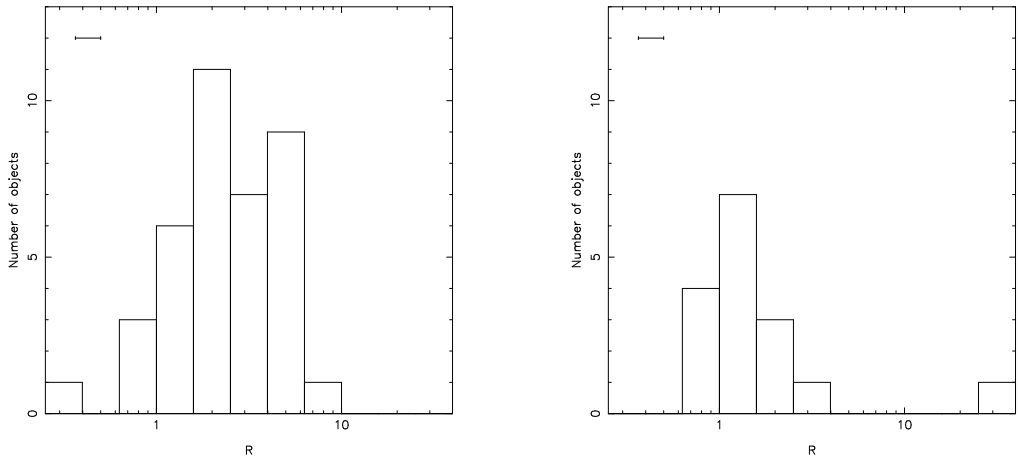}{f3b.eps}
\caption[Distribution of $R$ for the lobe sample]{Distribution of $R$
  values for the lobe sample. The left-hand plot shows the detected
  lobes and the right-hand plot shows the upper limits for
  non-detected lobes. A representative error bar is shown in the top
  left-hand corner of the left-hand plot.}
\label{lobehist}
\end{center}
\end{figure}

To study the overall properties of the sample, we first constructed a
histogram of $R$, the ratio of observed to predicted X-ray flux at
equipartition. Fig.~\ref{lobehist} shows histograms of $R$ for the
detected and non-detected lobes. Note that $R = 1$ means that the CMB
plus SSC model with an equipartition magnetic field and filling factor
of unity can explain the observed X-ray flux. For $R > 1$, in an IC
model, either the magnetic field is lower than the equipartition
value, i.e., the lobes are electron dominated, or an additional photon
field is present; $R < 1$ implies magnetic domination. We neglect here
the effects of a filling factor less than one, which could be the case
either for electrons, magnetic field or both. If the electrons fill
only a fraction of the lobe volume with a uniform field, we will {\it
underestimate} $R$, since the predicted CMB IC flux depends on the
number density of electrons, which we will have overestimated. If
there are strong magnetic field variations, but a uniform electron
density, we will {\it overestimate} $R$, because our prediction for
the number density and therefore CMB IC flux will be an
underestimation. The effect of filling factor is discussed in more
detail in \citet{hw00a}.

Table~\ref{bfield} gives the measured and equipartition magnetic field
strengths or upper limits and their ratio. We also list the ratio of
electron to magnetic field energy densities, for comparison with other
work in the literature. However, the electron and magnetic field
energies are sensitive to small changes in magnetic field strength, so
that the uncertainty on the value of $U_{e}/U_{b}$ is large. We
therefore consider $R$ and $B_{obs}/B_{eq}$ to be better measures of
the departure from equipartition. For comparison with other results in
the literature, we note that $R$ relates to the two other commonly
used measures of the departure from equipartition as $(B_{obs}/B_{eq})
\propto R^{\frac{-2}{\delta+1}}$ and $(U_{e}/U_{b}) \propto
R^{\frac{\delta+5}{\delta+1}}$, where $\delta$ is the electron energy
index ($ 2 \le \delta \le 3$).

\clearpage

\clearpage

The distribution of $R$ values for the detected sources is quite
narrow, with the most extreme values being 0.3 and 7.3 ($U_{e}/U_{b}$
ranges from 0.2 to 53). The majority of the sources have $R > 1$, and
appear to be distributed around a peak at $R \sim 2$. However, the
upper limits, in addition to the one detected source with $R < 1$,
show that some FRII radio lobes could be magnetically dominated by at
least a factor of 2 (or have a strongly structured magnetic field).
Since the non-detections are only a small fraction of the sample, we
can conclude that $>36/54$ lobes, or $\sim 70$ percent of FRII radio
galaxies and quasars if our sample is representative, are either at
equipartition or electron dominated.

\begin{figure}
\begin{center}
\plotone{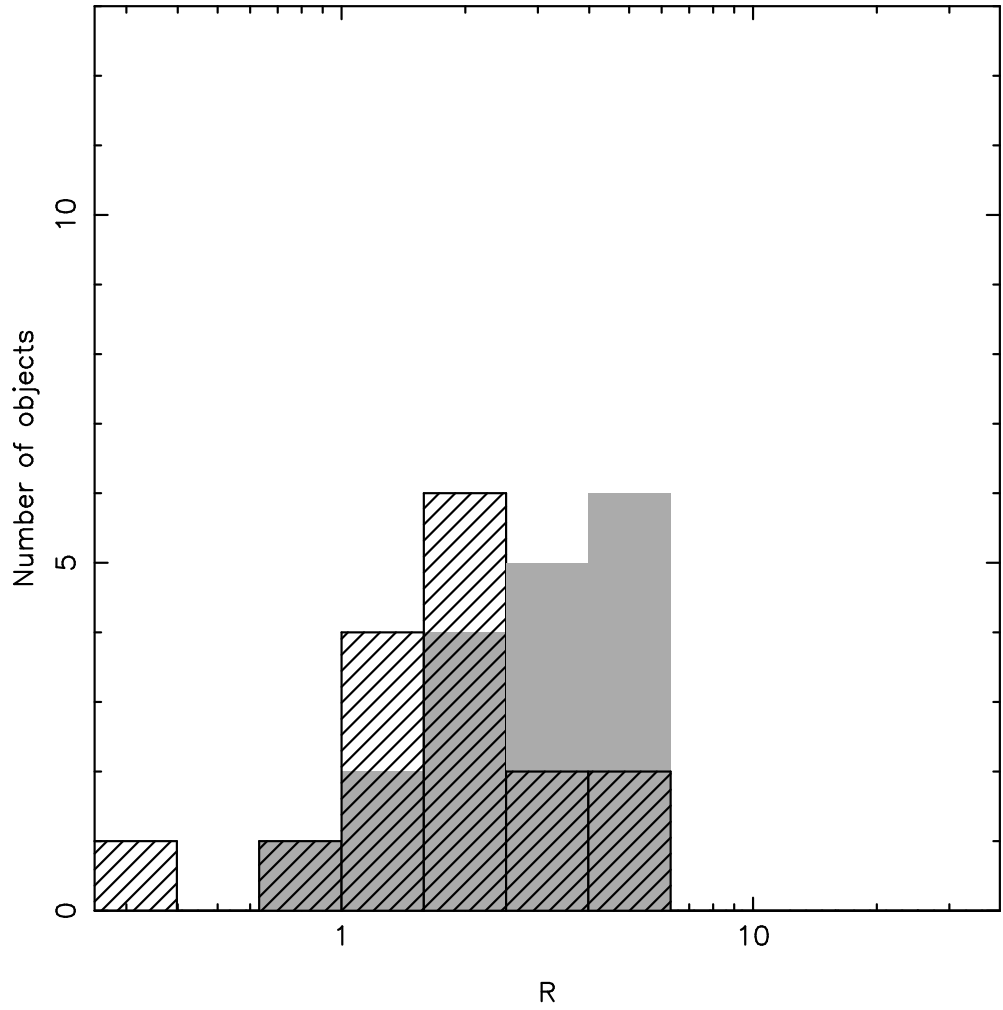}
\caption[$R$ distribution for narrow-line and broad-line objects]{The
  $R$ distribution for the narrow-line and broad-line objects.
  The distribution for broad-line radio galaxies and quasars is
  indicated with filled rectangles, and that for narrow-line radio
  galaxies is overplotted in hatched rectangles.}
\label{types}
\end{center}
\end{figure}

We next examined whether the type of radio source affects the observed
$R$ value, by comparing the distributions of $R$ for narrow-line radio
galaxies and for broad-line objects (broad-line radio galaxies and
radio-loud quasars). In the widely accepted unification model for
radio galaxies and radio-loud quasars \citep[e.g.,][]{bar89,up95},
these two categories of source, which possess different optical
properties but similar radio structure, are thought to be the same
objects seen at different angles to the line of sight. The narrow-line
objects are thought to occupy angles between $45^{\circ}$ and the
plane of the sky, whereas the radio-loud quasars and broad-line radio
galaxies occupy angles between $45^{\circ}$ and the line of sight,
with the difference in optical properties resulting from the presence
of a torus of cold material along our line of sight to the narrow-line
objects, which obscures the AGN in those sources. We excluded the
(few) low-excitation objects \citep{lai94} from our comparison to
avoid confusion. Fig.~\ref{types} shows a histogram with the two
categories of source indicated (only detected lobes are included;
however, the fraction of non-detections in the two subsamples are
similar). It is immediately apparent that the distributions of $R$
differ, in the sense that broad-line objects typically have higher
values of $R$. The broad-line objects nearly all have $R > 2$, whereas
the narrow-line objects mainly have $R < 2$. The median values of the
samples are 2.1 (narrow-line) and 3.1 (broad-line). A median test
rejects the hypothesis that the two subsamples have the same median
with $\sim 92$ percent probability.

\clearpage

\clearpage

One likely explanation for this marginally significant difference is
the effect of projection on the volumes of the lobes. The predicted
X-ray flux from CMB IC is proportional to the product of lobe volume
and electron density. The electron density scales as $V^{-4/7}$
\citep{h04b}, so that the predicted X-ray flux $S_{cmb} \propto
V^{3/7}$. Since we have not taken projection effects into account,
this means that for most sources we have underestimated the source
volume and therefore $S_{cmb}$, so that $R$ for a given source will be
likely to be overestimated. The effect will be at its most severe for
the broad-line radio galaxies and quasars, thought to be within 45
degrees of the line-of-sight, where the volumes will have been
significantly underestimated. If we assume that the population of
narrow-line radio galaxies occupies all angles between $45^{\circ}$
and the plane of the sky with equal probability, then the most
probable angle at which to observe a narrow-line radio galaxy is at
$\sim 70$ degrees, where $R$ will be overestimated by a factor of
$\sim 1.06$, assuming that volume scales as $l$, where $l$ is the
observed lobe length. Similarly, assuming that the population of
broad-line radio galaxies and quasars occupies all angles between
$45^{\circ}$ and the line of sight with equal probability, then the
most probable angle at which to observe a broad-line radio galaxy or
quasar is at $\sim 30$ degrees, where $R$ will be overestimated by a
factor of $\sim 1.34$. (Note that for angles less than 5 -- 10
degrees, $R$ can be overestimated by a factor of $> 2$). These results
show that the difference in the medians of the two samples cannot
entirely be explained by a model in which the intrinsic value of $R$
is the same for all radio galaxies and quasars.

To investigate this further, we carried out Monte Carlo simulations to
examine whether a narrow distribution of intrinsic $R$ values could
produce the observed distribution in $R$ as a result of projection
effects. We simulated samples of $10^{6}$ radio galaxies and quasars,
distributed at angles to the line of sight with a probability
distribution $P(\theta){\rm d}\theta = \sin(\theta){\rm d}\theta$
(i.e., based on the assumption that the lobes are randomly oriented
with respect to the plane of the sky), and having an intrinsic
Gaussian distribution of $R$ with a mean $<R>$ and variance $\sigma$.
We then determined the {\it observed} $R$ for each simulated source,
taking into account projection [$R_{app} \propto
(\sin(\theta))^{-3/7}$], assuming cylindrical lobes (with $V \propto
l$). We compared the simulated distribution of $R_{app}$ to the
observed distribution using a K-S test, and found that the intrinsic
values for $<R>$ and $\sigma$ that give the best match to the observed
data are 2.5 and 1.15, respectively. We next tested whether the
observed distributions of $R$ for the narrow-line and broad-line
objects could separately be explained by this intrinsic distribution.
We find that the broad-line objects have a $\sim 40$ percent
probability of being drawn from a parent population having this
intrinsic distribution; however, the narrow-line objects have only a
$\sim 3$ percent chance of being drawn from such a population. The
intrinsic distribution that gives the best fit to the narrow-line
objects alone has $<R> = 1.7$ and $\sigma=1.25$ (observed $R$ for the
narrow-line and broad-line subsamples are $2.1\pm1.1$ and $3.2\pm1.0$,
respectively).

We therefore conclude that projection is likely to be important in
explaining the distribution of observed $R$ values; however, some
additional explanation may be needed to explain the differences
between the narrow-line and broad-line objects. We note, however, that
the {\it actual} distribution of inclination angles in this sample is
unknown; it is clear from the fact that it contains roughly equal
numbers of narrow and broad-line objects that the probability
distribution of line-of-sight angle we used for the simulation does
not accurately represent the actual distribution in our sample. It is
also possible that the high redshift quasars (which often have high
$R$ values) may be biased towards small angles to the line of sight,
since the radio flux observed at 178-MHz in the highest redshift
objects may have contributions from beamed components (although this
will also reduce the intrinsic radio flux, which will act in the
opposite direction to raise the observed value of $R$). Another
possibility is that the systematic uncertainties in the X-ray analysis
may be worse for the quasars, as contamination from the bright central
source will be harder to remove; this could be compounded by the fact
that the quasars are typically more distant and so have smaller
angular sizes (although we find no correlation between $R$ and angular
size).

\clearpage

\begin{deluxetable}{lrrrr}
\tablecaption{Magnetic field
strengths and contributions to energy density.}
\tablewidth{10cm}
\tablehead{Source&$B_{\rm eq}$ (nT)&$B_{\rm obs}$ (nT)&$B_{\rm
    obs}/B_{\rm eq}$&$U_{\rm e}/U_{B}$}
\startdata
3C\,6.1N&3.5&$>3.5$&$>1.0$&$<1$\\
3C\,6.1S&3.4&$>2.5$&$>0.74$&$<3$\\
3C\,9W&10.0&11.0&1.1&0.70\\
3C\,47N&1.2&0.6&0.5&15\\
3C\,47S&1.1&0.4&0.36&53\\
3C\,109N&1.3&0.85&0.71&5\\
3C\,109S&1.3&0.85&0.71&5\\
3C\,173.1N&1.8&1.2&0.67&4\\
3C\,173.1S&1.3&$>1.0$&$>0.77$&$<3$\\
3C\,179E&2.9&1.3&0.45&$<19$\\
3C\,179W&2.9&1.1&0.38&37\\
3C\,200&1.4&1.1&0.79&3\\
3C\,207W&3.2&2.0&0.63&6\\
3C\,212S&7.4&$>1.0$&$>0.14$&$<177$\\
3C\,215N&1.2&0.6&0.5&13\\
3C\,215S&1.0&0.45&0.45&25\\
3C\,219N&0.6&0.35&0.58&6\\
3C\,219S&0.7&0.4&0.57&8\\
3C\,220.1E&1.7&$>1.7$&$>1$&$<1$\\
3C\,220.1W&1.9&$>1.9$&$>1$&$<1$\\
3C\,223N&0.35&0.22&0.63&5\\
3C\,223S&0.37&0.20&0.54&9\\
3C\,228N&2.3&$>1.8$&$>0.78$&$<2$\\
3C\,228S&2.4&$>2.4$&$>1$&$<1$\\
3C\,254W&3.9&$>3.9$&$>1.0$&$<1$\\
3C\,265E&2.1&1.3&0.62&7\\
3C\,265W&2.3&2.3&1.0&1\\
3C\,275.1N&2.5&$>2.8$&$>1.1$&$<0.7$\\
3C\,275.1S&3.1&2.8&0.9&1\\
3C\,280&5.5&9.0&1.6&0.2\\
3C\,281N&1.4&0.9&0.64&6\\
3C\,281S&1.4&$>1.0$&$>0.71$&$<3$\\
3C\,284E&0.52&0.40&0.76&3\\
3C\,284W&0.48&0.48&1.0&1\\
3C\,303E&1.9&$>0.9$&$>0.47$&$<15$\\
3C\,321W&0.60&$>0.7$&$>1.2$&$<0.6$\\
3C\,334N&1.5&1.5&1&1\\
3C\,334S&1.8&1.1&0.61&6\\
3C\,390.3N&0.7&$>0.7$&$>1$&$<1$\\
3C\,390.3S&0.7&$>0.7$&$>1$&$<1$\\
3C\,403N&0.5&0.2&0.40&27\\
3C\,403S&0.5&0.2&0.40&23\\
3C\,427.1N&3.3&$>3.3$&$>1$&$<1$\\
3C\,427.1S&3.7&3.7&1&1\\
3C\,452&0.5&0.25&0.50&9\\
\enddata
\label{bfield}
\tablecomments{Column 2 is the equipartition magnetic field strength,
column 3 is the magnetic field strength inferred from the level of
X-ray flux, column 4, the ratio of observed to equipartition field
strength, and column 5 the ratio of electron energy density to
magnetic field energy density. Note that 1nT = 10$\mu$G.}
\end{deluxetable}

\clearpage

\citet{h04b} carried out a similar analysis to that presented here for
the X-ray properties of hotspots in FRII radio sources. They found
that hotspots exhibit a large range in $R$ values, up to $R \sim
1000$, and conclude that a second X-ray emission component due to
synchrotron radiation must be present in some hotspots. In this
analysis we find that the IC model can explain all X-ray lobe
detections, with magnetic field strengths ranging from a fifth of the
equipartition value to slightly higher than the equipartition value.
This is unsurprising, as there is no known efficient acceleration
mechanism in the lobes that can produce electrons of
X-ray-synchrotron-emitting energies, whereas electrons at the hotspots
could be shock-accelerated to the energies required for X-ray
emission. Hardcastle et al. also find that $R$ is correlated with
radio luminosity, so that the highest $R$ values are found in the
weakest radio sources; they interpret this as being caused by a
luminosity-dependent cutoff in the maximum energy to which the
electrons are accelerated at the hotspot. We find no correlation
between $R$ and radio luminosity for the radio lobe sample. There is
also no correlation between $R$ and redshift. Finally, there is no
correlation between the lobe $R$ values and the hotspot $R$ values for
the same sources taken from Hardcastle et al., as expected, since we
believe that the X-ray emission mechanisms in lobes and hotspots are
different.

As part of a study of X-ray emission from jets, hotspots and lobes,
\citet{kat04} investigated the X-ray emission processes in a sample of
18 previously detected radio lobes. They conclude that inverse-Compton
emission with an equipartition magnetic field is the best model for
lobe X-ray emission for their smaller sample, in good agreement with
our results. Together with the work we present here, these results
provide strong support for the argument that FRII radio lobes are near
to equipartition.

To summarize, we find that detectable X-ray emission from the lobes of
FRII radio galaxies and quasars is common, that it is due
predominantly to IC scattering of CMB photons, and that most FRII
sources are close to equipartition, with the energy densities perhaps
being electron-dominated by a factor of a few. In the next sections,
we discuss reliability issues, and alternative explanations of our
results.

\section{Discussion}

\subsection{Reliability issues}

As discussed in Appendix~\ref{comp}, our conclusions in several cases
differ significantly from the work of previous authors. It is
extremely difficult to obtain a correct flux measurement for
lobe-related X-ray emission, because the best choice of extraction
regions is sometimes uncertain. It is difficult to avoid AGN
contamination, particularly when the radio-lobe emission lies close to
the core, and it is also important to exclude any contribution from a
hot-gas atmosphere. We have carefully chosen our background regions to
be at the same distance from the nucleus as the source regions,
although in two cases (3C\,200 and 3C\,452) this was not possible
because the radio-related X-ray emission surrounds the core. We
believe that it is the difficulty of correctly separating the
different components of the X-ray emission that has led to discrepant
results in the literature. It is extremely unlikely that any of our
flux measurements are {\it under}-estimates of the lobe IC emission,
so that any systematic uncertainty in the $R$ values is likely to be
in the direction of overestimation. As mentioned in
Section~\ref{sec:stat}, overestimation of $R$ values may be a
particular problem for some of the quasars with strong AGN and small
angular sizes.

\subsection{Assumptions about the low-energy electron population}
\label{sec:lowfreq}

As mentioned in Section~\ref{intro}, the properties of the low-energy
electron population in the radio lobes are not well constrained,
largely due to the lack of instruments capable of measuring the
sycnhrotron emission from this population. This is particularly
problematic when studying the IC/CMB process, as it is the low-energy
electrons that scatter the CMB to X-ray energies. Since we cannot use
observations to constrain directly the electron energy distribution
below 178 MHz for our sample, it is necessary to assume a
low-frequency spectral index, $\delta$, and cutoff energy,
$\gamma_{{\rm min}}$. For this work, we assumed that at low energies,
the electron population has an energy index $\delta = 2$, which
corresponds to the prediction from shock acceleration. This prediction
is supported by observations of hotspots \citep[e.g.,][]{mei97}, which
have the low-frequency spectral index predicted by the models. We
therefore assume that the electron population in the lobes has been
shock-accelerated while passing through the hotspots, resulting in an
initial energy distribution with $\delta = 2$ (corresponding to a
spectral index $\alpha = 0.5$). Spectral ageing has then steepened the
spectrum at observable frequencies to the observed values of 0.5 --
1.0. Our modelled energy distribution (Section~\ref{sec:synchmod}) is
therefore a power-law of index 2 at low energies, with a break to a
steeper slope in the observable radio region. We also chose to use
$\gamma_{{\rm min}}=10$, motivated partly by observations of
$\gamma_{{\rm min}} \sim 100 - 1000$ in hotspots
\citep[e.g.,][]{car91} -- we would expect a lower $\gamma_{{\rm min}}$
in lobes due to the effects of adiabatic expansion -- and partly to be
conservative. We believe that, given the lack of knowledge about this
electron population, our chosen electron energy distribution is
physically plausible; however, it is important to consider the effects
of varying $\delta$ and $\gamma_{{\rm min}}$, particularly since the
measured X-ray spectral index for those sources where a spectral model
could be fitted is steeper than 0.5 in a few cases.

We first tested the effects of varying $\gamma_{{\rm min}}$ for
several of our sources, covering a range of $R$ values. If we adopt
$\gamma_{{\rm min}} = 1000$, corresponding to the lower limit of
observed radio emission from lobes, we find new $R$ values that do not
differ from the quoted values (Table~\ref{icresults}) for
$\gamma_{{\rm min}} = 10$ within the 1$\sigma$ errors. The reason that
the prediction for X-ray inverse-Compton emission does not change
significantly is that, while the reduced energy range decreases the
electron density, the normalization of the electron energy spectrum
increases in order to maintain equipartition.

We next tested the effect of varying $\delta$. An alternative approach
to our method is to assume that the electron energy index implied by
the low-frequency radio spectral index can be extrapolated back to
$\gamma_{{\rm min}}$. We tested the effect of this assumption, using
the 3CRR spectral index $\alpha$ measured between 178 MHz and 750 MHz
(tabulated in Table~\ref{lobe_sample}), which are always greater than
0.5, for several of our sources, using $\gamma_{{\rm min}}=10$ and
assuming $\delta = 1+2\alpha$ and including a break or high-energy
cutoff in the spectrum as needed to fit to the radio spectrum, as in
our main analysis. We found that for all of the sources this resulted
in a {\it lower} prediction for the X-ray inverse-Compton flux,
increasing the $R$ values by a factor of $\sim 2$. The reason for the
lower IC prediction in this case is that the equipartition requirement
causes the electron energy spectrum normalization to be lowered with
respect to the $\delta = 2$ calculation because of the large
contribution to the electron energy density made by the additional
low-energy electrons. If we adopt $\gamma_{{\rm min}} = 1000$ for this
analysis, the electron normalization increases again, because of the
reduced energy range, so that the resulting $R$ values are again
roughly consistent with those in Table~\ref{icresults}. We therefore
conclude that uncertainty in the distribution of electrons at low
energies introduces at most a factor of 2 uncertainty into our quoted
$R$ values. This corresponds to a factor of 0.7 in $B_{obs}/B_{eq}$
and 4 in $U_{e}/U_{b}$.

\subsection{Anisotropic inverse-Compton emission from a
  nuclear photon field}

Our analysis for 3C\,265 in Section~\ref{sec:3c265} and for 3C\,284
\citep{c04}, as well as the analysis of \cite{bel04} for 3C\,184,
shows that IC scattering of the photon field from a hidden quasar does
not appear to be the dominant X-ray emission mechanism in these
sources. As mentioned in Section~\ref{sec:3c265}, the X-ray emission
from the IC/nuclear process would be brightest towards the nucleus and
decrease rapidly with radius; this is not the morphology that is
observed. In contrast, the X-ray emission from CMB IC is expected to
follow closely the structure of the low-frequency radio emission,
which appears to be the case for those sources where the
signal-to-noise is sufficiently high to observe the spatial
distribution of emission. Our calculations for IC/nuclear
(Section~\ref{sec:3c265}), in contrast to the work of other authors,
assume that the incident photon field is emitting isotropically. The
justification for this assumption comes from observations of the
infrared emission from narrow-line and broad-line objects which show
that the infrared properties of the two types of object are the same
\citep{mei01}. IC/nuclear emission may be important in some sources in
this sample, and may help explain the different distributions of
narrow-line and broad-line objects. 3C\,207 \citep{brun02} may be an
example where this process is important. However, we have shown that
CMB IC with a near equipartition magnetic field can account for the
majority of the observed X-ray emission in most narrow-line radio
galaxies (and, if projection is taken into account, in many broad-line
and objects as well); we conclude that in most cases IC/nuclear is not the
dominant process.

\subsection{An alternative interpretation: shock-heated gas?}

As discussed in the context of our observations of 3C\,223 and
3C\,284 \citep{c04}, it is also possible that there is hot, shocked gas
surrounding the radio lobes of some sources. The emission from such
gas could be mistaken for IC emission, as it is difficult to
distinguish spectrally between these models due to the small number of
counts from most of the lobes. However, in the cases where spectral
fitting could be performed, a thermal model was usually a
significantly poorer fit to the data. As mentioned above, the lack of
a correlation between $R$ and redshift also suggests that the emission
is not dominated by such shocked gas, which would be difficult to
detect at high redshifts (based on the assumption that most FRIIs reside in
groups, as found by \cite{bes04}, and assuming a typical group
luminosity of $10^{42}$ erg s$^{-1}$). In addition, many of the
sources have radio morphologies similar to 3C\,223 and 3C\,284
\citep{c04}, for which we argue that highly supersonic
expansion is unlikely. These arguments do not rule out some
contribution from hot gas. However, the results from 3C\,452, where a
two-component model could be fitted, show that it is not possible to
explain all of the excess X-ray emission above the equipartition
prediction by contamination from thermal emission.

As an additional test, we can compare the expected luminosity of
shock-heated gas with the observed luminosity in sources where some
estimate of the physical conditions in the external medium has been
made. We assume a shock-heated shell surrounding the entire lobe. The
temperature of the shock-heated gas is unknown, though the evidence
from spectral fits, where these are possible, is that it must be high
($\ga 5$ keV). Fortunately {\it Chandra}'s response to gas hotter than
a few keV is only weakly sensitive to temperature, so this does not
restrict our ability to carry out these calculations. We make the
assumption that the shock-heated material is gas swept up in the radio
lobe's expansion, compressed by some compression factor $k$: then if
the number of particles swept up by the lobe is $N$, and the lobe's
volume is $V$, the mean density of particles in the shell is $kN/V$.
Assuming uniform density, the luminosity from the shocked shell is
$CN^2k/V$, where $C$ is a constant (depending on the luminosity band
of interest, the metal abundance of the shocked material, and, weakly,
on its temperature). The compression factor is unknown, but given the
close match of the detected X-rays to the shape of the radio-emitting
lobes, must be significantly greater than 1: application of standard
jump conditions would give $k=4$, while the observed shock around the
southern lobe of Cen A \citep{kra03} corresponds to $k \approx 10$. We
have calculated the expected luminosity for shocked shells in several
sources for which we have estimates of the group/cluster parameters
\citep[e.g.,][]{h02a,c04}. In general we find that the expected
luminosity for $k>1$ exceeds the observed luminosity of the lobes;
what we observe is too faint to be compressed, swept-up gas. (The
exceptions to this rule are sources that are found to lie in
reasonably rich environments: for example, $k=4$ is allowed by the
data for 3C\,263.) While the results are uncertain because the
physical parameters of the environments are poorly constrained, we
consider the general incompatibility of this simple model with the
observations to be an additional argument against the picture in which
the lobe-related X-rays are due to shock-heated thermal material. If
supersonic expansion occurred in a small region of the lobe, e.g.,
around a hotspot, then the expected luminosity from the shocked gas
would be lower, and could be compatible with the observations.
However, in the largest sources with high signal-to-noise detections
(e.g., 3C\,452), the X-ray emission is not localized in this way, and
any emission from shock-heated gas close to hotspots will have been
excluded from our analysis. We therefore conclude that this scenario
is probably unimportant for the majority of sources in the sample,
although it could contribute in some high $R$ sources where the data
quality is insufficient to rule out localized shock heating.

\subsection{Implications for particle content}

If the lobes of FRII sources contained an energetically dominant
population of relativistic protons (with a high ratio of $U_{\rm
p}/U_{\rm e}$), and the energy densities in magnetic field and {\it
particles} were similar, then $R$ would be expected to be typically
less than unity. Our results therefore rule out a model where FRII
radio lobes have an energetically important proton population and are
at equipartition. It is not possible to rule out directly a model in
which radio lobes are highly particle-dominated, i.e., where there is
an energetically important population of protons giving a total energy
density in particles that is an order of magnitude or more higher than
that in the magnetic field. However, such a model cannot explain why
the measured magnetic field strengths are always close to the value
for equipartition between relativistic {\it electrons} and magnetic
field, unless the mechanism for achieving equipartition requires
timescales longer than the lobe lifetimes for protons but not
electrons. The results of our survey of FRII radio lobes support the
conclusions of our earlier papers on smaller samples of sources
\citep{h02a,c04}: the presence of an energetically dominant population
of protons is unlikely, because it requires that the magnetic field
energy density tends to be similar to the electron energy density
rather than the conjectured overall energy density in relativistic
particles.

\section{Conclusions}

Our study of the X-ray emission from the lobes of FRII radio galaxies
and quasars has shown that they can be magnetically dominated by at
least a factor of 2; however, $>70$ percent of the sample are at
equipartition or electron dominated. There is a reasonably narrow
distribution of $R$ values, where $R$ is the ratio of observed to
predicted emission from CMB IC from synchrotron-emitting electrons at
equipartition. The distribution peaks at $R \sim 2$, which corresponds
to magnetic field strengths within 35 percent of the equipartition
value, or electron dominance ($U_{e}/U_{b}$) by a factor of $\sim 5$.
That the distribution is narrow and close to the expectation for
equipartition between relativistic electrons and magnetic field shows
that an energetically dominant proton population in FRII radio sources
is unlikely. The distribution of apparent $R$ values differs for
narrow-line radio galaxies and broad-line objects (broad-line radio
galaxies and quasars); this is due in part to projection effects, but
may also be caused by worse systematic uncertainties for more distant
objects. We argue that IC scattering of infrared and optical photons
from the nucleus is unlikely to be the dominant X-ray emission process
in the majority of radio galaxies and quasars, although it may play
some r\^ole in smaller objects.

\clearpage

\acknowledgments

We thank the referee for helpful comments. We gratefully acknowledge
support from PPARC (studentship for JHC and research assistantship for
EB) and the Royal Society (research fellowship for MJH). This work was
partially supported by NASA grant GO3-4132X.

The National Radio Astronomy Observatory is a facility of the National
Science Foundation operated under cooperative agreement by Associated
Universities, Inc.

\appendix
\section{Comparison of results for sources previously presented by other authors}
\label{comp}

In this section, we briefly discuss the lobe emission from the
individual sources for which previously published results exist and
compare our results with those of other authors.

\subsection{3C\,9}

\citet{fcj03} discuss the X-ray emission from this source, attributing
it to inverse-Compton scattering of the CMB; however, they do not
carry out a detailed calculation, and do not separate the X-ray
emission from the eastern radio components (likely to be jet-related)
and the western lobe, which we analysed above. Our results are in rough
agreement with their less detailed analysis.

\subsection{3C\,179}

\citet{sam02} discuss the {\it Chandra} observation of 3C\,179 in the
context of a survey of jets studied with {\it HST} and {\it Chandra}.
Their Fig.~1 shows a smoothed image of the X-ray emission; however,
their choice of greyscale means that the lobe-related emission does
not show up in the image, which emphasizes the jet and hotspot
emission. Our 0.5 -- 5.0 keV image (Fig.~\ref{lobes1}) excludes higher
energy background counts included in their image (which used a larger
energy range), and shows clearly an excess of counts associated with
both lobes. Sambruna et al. do not mention the presence of this
emission.

\subsection{3C\,207}

The {\it Chandra} observation of 3C\,207 was presented by
\citet{brun02}. We measured only 25 0.5-5 keV counts using background
subtraction at the same distance from the nucleus, whereas they were
able to fit a spectrum with 9 bins. In their discussion of background
subtraction they argue that the choice of background region is not
important as the {\it Chandra} background is extremely low; this
suggests that they have not subtracted off the component of background
from the wings of the PSF, which is important, as can be seen from
their smoothed image. They list several reasons why they believe that
IC scattering of the CMB is not the dominant process. They argue that
no emission is seen from the western lobe, which is of similar radio
luminosity and size to the eastern one; however, as the source is of
small angular size, and the western lobe is dominated by jet emission,
it is not possible to obtain a strong upper limit on its lobe-related
X-ray emission, particularly in the presence of high background due to
AGN emission. Brunetti et al. argue that the X-ray spectral index is
flatter than the radio spectral index; however, firstly there do not
appear to be sufficient counts to constrain the spectral index, and
secondly they consider only the high frequency spectral index, whereas
it is lower energy electrons that will scatter the CMB and the nuclear
emission. We obtained an $R$ value of 3 for the western lobe of
3C\,207 (the intrinsic value may be lower due to projection, see
above). It is therefore possible that some contribution to the X-ray
flux comes from the IC/nuclear process as argued by \citet{brun02},
but it appears that CMB IC can explain a significant fraction of the
observed emission if the lobe is near equipartition, or all of the
X-rays if the lobe is modestly electron-dominated or at a small angle
to the line of sight.

\subsection{3C\,219}

\citet{com03} present the {\it Chandra} observation of 3C\,219 and
attribute the lobe-related emission to inverse-Compton scattering of
the CMB and nuclear AGN photons. Our spectral results for the northern
lobe are consistent with the values they obtain for the entire source.
However, their quoted flux is approximately twice our measured total
flux from both lobes. This is probably due to their choice of spectral
extraction region, which does not exclude the jet or nothern hotspot
regions. Additionally, they do not specify their choice of background
region: if it is off-axis, then their spectrum could contain
significant AGN contamination. We obtain a comparable flux to their
measured value if we use a large elliptical extraction region that
includes the jet and northern hotspot, and use an off-source
background region. We believe that our choice of extraction regions is
preferable, as our regions follow the radio structure more closely,
and because our background region, at the same distance from the core
as the source regions, will remove contamination from the AGN and
hot-gas environment. Therefore, we disagree with their conclusions
that the lobes are electron-dominated by up to a factor of 100, and
find that this factor is more than an order of magnitude lower.

\subsection{3C\,265}
\label{sec:3c265}

\citet{bon04} present the results of an analysis of the {\it Chandra}
observation of 3C\,265. They interpret the origin of the X-ray
emission as IC scattering of AGN photons; however, our results show
that CMB and SSC IC emission in an equipartition field can account for
$>1/3$ of the observed X-ray emission. It is nevertheless possible
that IC/nuclear scattering makes a significant contribution to the
flux from the eastern lobe, thought to be pointing away from us. To
test this, we carried out similar calculations for this source to
those for the eastern lobe of 3C\,284 presented by \cite{c04}.

To model the illuminating flux, we used the infrared measurements (at
4.5, 6.7 and 12.0 $\mu$m) for 3C\,265 of \citet{sie04}, extrapolating
to a frequency range of $4 \times 10^{12}$ to $10^{15}$ Hz. In order
to model the shape of the spectrum at lower frequencies, we scaled the
lower frequency spectrum for 3C\,295 \citep{mei01} to a value
appropriate for the normalization of 3C\,265's infrared spectrum. We
used the same parameters for the electron energy spectrum as in the
analysis of Section~\ref{sec:synchmod}, with $\gamma_{{\rm min}}$ of
10, so that the predicted flux is an upper limit on the contribution
from IC/nuclear. We find that the contribution to the predicted 1-keV
flux density from IC scattering of this photon field to be less than
0.03 nJy for all choices of inclination angle. Therefore, if the
infrared emission is isotropic, this process cannot account for the
observed flux level. In order to produce the additional X-ray flux
(the observed excess of 1.7 nJy above the CMB IC prediction; see
Table~\ref{icresults}), the luminosity of the quasar as seen by the
lobes would have to be $\sim 60$ times more luminous than the
isotropic luminosity, so that a beam of small opening angle would be
required. This more luminous photon field, if symmetric on each side
of the source, would also produce a significant flux from the western
radio lobe (e.g., $\sim 1$ nJy for the most probable angle of
$69^{\circ}$), so that IC/nuclear scattering in an equipartition
magnetic field cannot explain the observed fluxes. Another argument
against illumination from a narrow cone of infrared emission is the
lack of evidence for differences in the infrared properties of
narrow-line and broad line radio galaxies \citep{mei01}. Finally, the
IC/nuclear model predicts a steep decrease in the X-ray flux with
distance from the nucleus; this is not seen in the X-ray data. The
asymmetry in X-ray-to-radio flux is therefore more plausibly explained
by a difference in the relative magnetic field strength of the two
lobes, perhaps due to their different sizes.

There are several differences between our method and that of Bondi et
al., which may explain our different conclusions as to the dominant
photon population. Our X-ray observational measurements appear to be
in reasonable agreement; however, from our L-band radio map we measure
a much smaller flux ratio between the two lobes than that quoted by
Bondi et al. (a ratio of 2.7 between the east and west lobes). We also
use a concordance model cosmology, whereas Bondi et al. use a value of
$q_0 = 0.5$; this has a significant effect on the volumes and the CMB
energy density. The different radio measurements and cosmology may
explain why our value for the equipartition magnetic field (2.4 nT) is
a factor of two below theirs. We used an illuminating spectrum based
directly on the infrared measurements of \citet{sie04} for 3C\,265,
which gives an integrated luminosity just below the lower limit of the
range quoted by Bondi et al. In contrast, they use a value close to
the upper limit of their quoted range. Finally, we assumed that the
illuminating source is isotropic, as supported by arguments from
unified models (see above). We believe our choice of cosmology is more
appropriate, and our parametrization of the illuminating source is
based directly on infrared measurements, and so we conclude that, in
contrast to the findings of Bondi et al., IC from CMB photons in a
near-equipartition field ($B \sim 0.6B_{eq}$) can explain the observed
X-ray flux, and that the contribution from IC/nuclear is not dominant
in this source.

\subsection{3C\,281}

\citet{cf03} presented the {\it Chandra} observation of 3C\,281 and
attributed the extended X-ray emission to the northern hotspot and
environment. They mention the presence of extended soft emission along
the jet axis of the four sources they observe in the context of lobe
IC emission; however, they concentrated primarily on the environmental
properties and did not carry out any analysis of radio-related
emission.  The emission they considered to be hotspot-related is not
sufficiently compact or directly associated with the radio hotspot,
leading \citet{h04b} to argue that the radio-related emission is
associated instead with the lobe.  Crawford \& Fabian found that the
luminosity of the extended emission is much lower than the values
estimated using {\it ROSAT}, which is probably due to the poorly known
{\it ROSAT} PSF. The true group or cluster luminosity may be even
lower, once the contribution from lobe-related emission is removed.

\subsection{3C\,452}

An in-depth study of the lobe-related emission from 3C\,452 was
presented by \citet{iso02}, including analysis of the spatial
structure of electrons and magnetic field in the lobes. They attribute
all of the non-thermal emission to IC scattering of the CMB and find a
best-fitting power-law plus Raymond-Smith model with similar spectral
parameters to those given in Table~\ref{lobespec}. They carry a simple
comparison of the ratios of synchrotron radiation to X-ray IC emission
and estimate that the lobes are electron dominated by a factor of
27$^{+25}_{-16}$. Their lower limit is roughly consistent with our
estimated value of 9, which was determined using a more detailed
synchrotron and inverse-Compton modelling procedure.

\end{document}